\documentclass[11pt]{article}

\usepackage{amsmath}
\usepackage{graphicx}
\usepackage{amsfonts}
\usepackage{amssymb}
\usepackage{epsfig}
\usepackage{color}
\usepackage{psfrag}
\usepackage{epstopdf}

\usepackage{caption}
\usepackage{subcaption}

\newcommand{\EQ}{\begin{equation}}
\newcommand{\EN}{\end{equation}}
\newcommand{\be}{\begin{equation}}
\newcommand{\ee}{\end{equation}}
\newcommand{\bea}{\begin{eqnarray}}
\newcommand{\eea}{\end{eqnarray}}

\DeclareMathOperator*{\SumInt}{%
\mathchoice%
  {\ooalign{$\displaystyle\sum$\cr\hidewidth$\displaystyle\int$\hidewidth\cr}}
  {\ooalign{\raisebox{.14\height}{\scalebox{.7}{$\textstyle\sum$}}\cr\hidewidth$\textstyle\int$\hidewidth\cr}}
  {\ooalign{\raisebox{.2\height}{\scalebox{.6}{$\scriptstyle\sum$}}\cr$\scriptstyle\int$\cr}}
  {\ooalign{\raisebox{.2\height}{\scalebox{.6}{$\scriptstyle\sum$}}\cr$\scriptstyle\int$\cr}}
}

\begin{document} \setcounter{page}{0}
\newpage
\setcounter{page}{0}
\renewcommand{\thefootnote}{\arabic{footnote}}
\newpage
\begin{titlepage}
\begin{flushright}
\end{flushright}
\vspace{0.5cm}
\begin{center}
{\large {\bf Quantum quenches from an excited state}}\\
\vspace{1.8cm}
{\large Gesualdo Delfino$^{1,2}$ and Marianna Sorba$^{1,2}$}\\
\vspace{0.5cm}
{\em $^1$SISSA -- Via Bonomea 265, 34136 Trieste, Italy}\\
{\em $^2$INFN sezione di Trieste, 34100 Trieste, Italy}\\
\end{center}
\vspace{1.2cm}

\renewcommand{\thefootnote}{\arabic{footnote}}
\setcounter{footnote}{0}

\begin{abstract}
\noindent
Determining the role of initial conditions in the late time evolution is a key issue for the theory of nonequilibrium dynamics of isolated quantum systems. Here we extend the theory of quantum quenches to the case in which before the quench the system is in an excited state. In particular, we show perturbatively in the size of the quench (and for arbitrarily strong interactions among the quasiparticles) that persistent oscillations of one-point functions require the presence of a one-quasiparticle contribution to the nonequilibrium state, as originally shown in [J.~Phys.~A~47~(2014)~402001] for the quenches from the ground state. Also in the present case, we argue that the results generically have nonperturbative implications. Oscillations staying undamped within the accessible time interval, far beyond the perturbative time scale, are nowadays observed in numerical simulations.
\end{abstract}
\end{titlepage}

\newpage
\tableofcontents

\section{Introduction}
The nonequilibrium dynamics of isolated quantum systems is characterized by the fact that expectation values are computed on nonequilibrium states which are superpositions of infinitely many eigenstates of the Hamiltonian, with different superpositions corresponding to different initializations of the dynamics. A key theoretical issue is to understand to which extent the late time evolution depends on the initial conditions. A first study for an infinite-dimensional space of initial conditions recently showed through which quantum mechanisms universal quantitative properties can emerge at large times \cite{q_int}. 
In the present paper we investigate the dependence on initial conditions for the case in which the nonequilibrium state is generated dynamically by a change of an interaction parameter at time $t=0$. This basic way of accessing nonequilibrium \cite{BMcD}, which has been called "quantum quench" \cite{SPS,CC} in analogy with thermal quenches of classical statistical systems, is itself quite nontrivial from the point of view of theoretical study. However, it was shown in \cite{quench} that for a homogeneous one-dimensional system with Hamiltonian
\EQ
H=\left\{
\begin{array}{l}
H_0\,,\hspace{3cm}t<0\,,\\
\\
H_0+\lambda\int_{-\infty}^\infty dx\,\Psi(x)\,,\hspace{.7cm}t>0\,,
\end{array}
\right.
\label{quench}
\EN
which for $t<0$ is in the ground state $|0\rangle$ of the pre-quench Hamiltonian $H_0$, it is possible to obtain {\it general} results perturbatively in the quench size $\lambda$, for the different quench operators $\Psi(x)$ and independently of the strength of the interaction among the quasiparticles. In particular, the result
\EQ
\langle \Phi(x,t)\rangle_0 =\langle\Phi\rangle_{\lambda}^{\textrm{eq}}+\lambda\left[\frac{2}{M^2}\,F_1^{\Psi} F_1^{\Phi}\, \cos Mt+O(t^{-3/2})\right]+O(\lambda^2)\,
\label{Phi_0}
\EN
is obtained for the one-point function\footnote{The subscript $0$ in (\ref{Phi_0}) refers to the number of quasiparticle excitations in the pre-quench state, which is zero for the ground state.}  of an operator $\Phi$ at large time \cite{quench}. Here $\langle\Phi\rangle_{\lambda}^{\textrm{eq}}$ is the equilibrium expectation value in the theory with the post-quench Hamiltonian \cite{DV}, $M$ the quasiparticle mass, and $F^{\cal O}_1$ the matrix element of ${\cal O}$ between the ground state $|0\rangle$ and the one-quasiparticle state\footnote{We refer for simplicity to the case of a single species of quasiparticles. Otherwise, the square bracket in (\ref{Phi_0}) is summed over species \cite{quench}.}. As a particularly interesting feature, the result (\ref{Phi_0}) first revealed that undamped oscillations of one-point functions are present whenever an internal symmetry does not cause the vanishing of the product  $F_1^{\Psi} F_1^{\Phi}$ of the one-quasiparticle matrix elements \cite{quench}. In particular, the undamped oscillations are absent in the case of noninteracting quasiparticles -- for which $F_1^{\Psi}=0$ -- thus showing that interaction makes a qualitative difference in nonequilibrium quantum dynamics.

The fact that the condition $F_1^{\Psi} F_1^{\Phi}\neq 0$ actually leads to the presence of undamped oscillations beyond the perturbative time scale was argued in \cite{oscill} along the lines that we will recall in section \ref{further} below. Remarkably, oscillations exhibiting no damping over hundreds of periods within the accessible time interval were then observed numerically\footnote{Undamped oscillations had been observed over much shorter time spans in previous numerical studies, see \cite{BCH,RMCKT,KCTC,Lukin,Liu} and the discussion in \cite{oscill}.} for a nonperturbative quench of the Ising spin chain in \cite{Jacopo}. 

The extension of the analytical results of \cite{quench} to higher spatial dimensions was given in \cite{oscillD}, where it was shown, in particular, that $F_1^{\Psi} F_1^{\Phi}\neq 0$ continues to be the condition for undamped oscillations in a homogeneous quench\footnote{For the more general case of inhomogeneous quenches there is an additional condition about the extensiveness of the quenched domain \cite{oscillD}, see also section \ref{further} below.}. Since $F^{\cal O}_1\neq 0$ means that ${\cal O}$ creates the quasiparticle, $F_1^{\Psi} F_1^{\Phi}\neq 0$ means that the post-quench theory possesses a neutral\footnote{Detailed illustrations of the role of symmetries and charges are given in \cite{DV,oscillD}.} quasiparticle ($F_1^{\Psi}\neq 0$) and that $\Phi$ couples to this quasiparticle ($F_1^{\Phi}\neq 0$). It is natural to ask to which extent the presence of persistent oscillations depends on the initial conditions. Answering this question, however, had proved too difficult so far. Indeed, changing the initial condition for the homogeneous quench (\ref{quench}) means starting with a pre-quench state other than the ground state of $H_0$, and then with a non-normalizable state whose direct use leads to undetermined expressions. 

Here we show how to overcome these technical difficulties considering the quench (\ref{quench}) when for $t<0$ the system is in the first excited state of its Hamiltonian $H_0$, which is a state with a single quasiparticle. The calculations are performed in presence of a regulator $R$ which prevents singularities and is eventually removed ($R\to\infty$) leaving a finite result for the observables. In this way we find that the large time behavior for the one-point function becomes
\EQ
\langle \Phi(x,t)\rangle_1 =E_\Phi+\lambda\left[\frac{6-4\sqrt{2}}{M^2}\,F_1^{\Psi} F_1^{\Phi}\, \cos Mt+O(t^{-1})\right]+O(\lambda^2)\,.
\label{Phi_1}
\EN 
We see that, although the amplitude of the undamped oscillations has changed, the condition for their presence continues to be $F_1^{\Psi} F_1^{\Phi}\neq 0$. The offset $E_\Phi$ in general differs at order $\lambda$ from $\langle\Phi\rangle_{\lambda}^{\textrm{eq}}$ in (\ref{Phi_0}). Actually, as we will see, $E_\Phi$ can differ from $\langle\Phi\rangle_{\lambda}^{\textrm{eq}}$ at order 1 for a quench in a spontaneously broken phase, for which the first excited state is a topological excitation.

The paper is organized as follows. In the next section we set up the quench from the excited state and obtain the general expression for the post-quench state and the one-point functions. The latter is then considered in the large time limit in section~3. Section~4 is devoted to the case of topological quasiparticles, while some final remarks are collected in the final section.

\section{Post-quench state and one-point functions}
We consider an infinite and homogeneous system in one spatial dimension which before the quench (i.e. for $t<0$) is in the first excited state, namely the one-quasiparticle state $|q\rangle$. The latter is an eigenstate of the pre-quench Hamiltonian $H_0$ with momentum $q$ and energy 
\EQ
E_q=\sqrt{M^2+q^2}\,,
\label{dr}
\EN
and is normalized as 
\EQ
\langle p|q\rangle=2\pi E_q\delta(p-q)\,.
\EN
A direct use of the pre-quench state $|q\rangle$ for the determination of  the post-quench dynamics, however, is complicated by the appearance of singular factors such as $\langle q|q\rangle$. These factors, appearing in the numerator and denominator of the expressions for the observables, lead to undeterminate expressions. It is then necessary to first prevent the singularities introducing a regulator to be removed at the end of the calculations. We do so starting with a pre-quench state in the form of the wave packet
\EQ
\int_{-\infty}^{\infty} dq\, f(q)\, |q\rangle\,,
\EN
with
\begin{equation}
f(q)= \frac{R}{\sqrt{2\pi}}\, e^{-\frac{R^2}{2}\, q^2}\,.
\end{equation}
The results corresponding to the pre-quench state\footnote{A state $|q_0\rangle$ can be obtained centering the Gaussian in $q_0$. Since we will only consider scalar operators, the results for $R\to\infty$ do not depend on $q_0$, and setting it to zero involves no loss of generality.} $|q=0\rangle$ are obtained taking the limit $R\to\infty$ in the final expressions for observable quantities.  

The state $|\psi_1\rangle$ produced by the quench is given by
\begin{equation}
|\psi_1\rangle=S_{\lambda}\int dq\, f(q)\, |q\rangle = T \exp\left(-i\lambda \int_0^{+\infty} dt \int_{-\infty}^{+\infty} dx\, \Psi(x,t) \right) \int dq\, f(q)\, |q\rangle\,,
\label{psi1_scattering}
\end{equation}
where $T$ denotes chronological order and $S_{\lambda}$ is the operator whose matrix elements $\langle n|S_{\lambda}|q\rangle$ give the probability amplitude that the quench induces the transition from $|q\rangle$ to $|n\rangle$. Here we adopt the compact notation $|n\rangle=|p_1,...,p_n\rangle$ for the $n$-quasiparticle states of the pre-quench theory, having energy and momentum 
\EQ
E=\sum_{i=1}^n E_{p_i}\,,\hspace{1cm}P=\sum_{i=1}^n p_i\,,
\EN
respectively. To first order in the quench parameter $\lambda$ we have
\begin{equation}
|\psi_1\rangle \simeq \int dq\, f(q)\, |q\rangle + 2\pi \lambda \SumInt\limits_{n,p_i} \int dq\, f(q)\, \delta (P -q)\,  \frac{[F^{\Psi}_{1,n}(q\,|\{p_i\})]^*}{E-E_q}\, |n\rangle\,,
\label{psi1}
\end{equation}
where we defined the matrix elements
\begin{equation}
F_{1,n}^{\cal O}(q\,|\{p_i\})= \langle q|{\cal O}(0,0)|p_1,...,p_n\rangle\,
\label{matrix_element}
\end{equation}
for a generic operator ${\cal O}(x,t)$, introduced the notation
\begin{equation}
\SumInt\limits_{n,p_i}=\sum_{n\neq 1} \frac{1}{n!} \int_{-\infty}^{+\infty} \prod_{i=1}^n \frac{dp_i}{2\pi E_{p_i}}\,,
\label{compact_SumInt}
\end{equation}
and used
\begin{equation}
{\cal O}(x,t)=e^{i\mathcal{P}x+iH_0 t}\,{\cal O}(0,0)\, e^{-i\mathcal{P}x -iH_0 t}\,,
\label{shift}
\end{equation}
with $\mathcal{P}$ the momentum operator. An infinitesimal imaginary part was given to the energy to make the time integral in (\ref{psi1_scattering}) convergent. The sum (\ref{compact_SumInt}) is taken on the non-negative integers with the exclusion of $n=1$, which in (\ref{psi1}) corresponds to $-iLT_+\lambda\int dq\, f(q)\,F_{1,1}^\Psi(q|q)$, where $LT_+$ is the infinite post-quench space-time volume; this contribution is due to mass renormalization and is subtracted by a corresponding counterterm in the Hamiltonian\footnote{See \cite{nonint} for an analogous subtraction in the equilibrium context.}. 

In an interacting theory, the matrix elements $F^{\Psi}_{1,n}$ entering (\ref{psi1}) are nonzero for any $n$, so that the state produced by the quench is a superposition of states containing any number of quasiparticles with all possible momenta. It is worth emphasizing that it is our ability to deal with this structure in its full complexity that allows us to obtain general analytical results about quantum quench dynamics.

The result (\ref{psi1}) gives in particular
\EQ
\langle \psi_1|\psi_1\rangle = \int dpdq\,f(p)f(q)\langle p|q\rangle\, +O(\lambda^2)\sim \sqrt{\pi}MR + O(\lambda^2)\,.
\label{norm}
\EN
Here and below the symbol $\sim$ indicates omission of terms subleading for $R\to\infty$.

The one-point function of a scalar Hermitian operator $\Phi(x,t)$ after a quench from the single-quasiparticle state is given by
\begin{align}
\langle \Phi(x,t)\rangle_1 & =\frac{\langle \psi_1|\Phi(x,t)|\psi_1\rangle}{\langle \psi_1|\psi_1\rangle} +D_{\Phi} \nonumber\\
&\sim\frac{1}{\langle \psi_1|\psi_1\rangle}\,\int dp\, dq\,f(p)f(q) \left[F_{1,1}^{\Phi}(p|q)\, \right.\nonumber\\
&+ \left. 2\pi\lambda \SumInt\limits_{n,p_i} \frac{\delta(P-q)}{E-E_P}\,2\text{Re}\left\{[F_{1,n}^{\Psi}(P\,|\{p_i\})]^* F_{1,n}^{\Phi}(p\,|\{p_i\})\,e^{i(E_p-E)t} \right\}\right]\nonumber\\
&+D_{\Phi}+ O(\lambda^2)\,.
\label{onepoint}
\end{align}
Here and below the limit $R\to\infty$ is understood. Then the limit $p,q\to 0$ is enforced by the Gaussian $f(p)f(q)$, so that the factors $e^{i(p-q)x}$ and $e^{i(E_p-E_q)t}$ produced by (\ref{shift}) can be omitted at leading order for large $R$; this yields the $x$-independence expected for the homogeneous system, as well as the $t$-independence of the pre-quench value
\EQ
A_\Phi=\frac{\int dp\, dq\,f(p)f(q)\,F_{1,1}^{\Phi}(p|q)}{\int dp\, dq\,f(p)f(q)\langle p|q\rangle}\,.
\label{prequench}
\EN
The term $D_{\Phi}$ is added to ensure continuity at $t=0$, namely to impose that $\langle \Phi(x,0)\rangle_1$ is equal to (\ref{prequench}). Defining
\EQ
B_\Phi(t)=\frac{2\pi}{\langle \psi_1|\psi_1\rangle}\,\SumInt\limits_{n,p_i}\,\int dp\, \frac{f(p)f(P)}{E-E_P}\,2\text{Re}\left\{[F_{1,n}^{\Psi}(P\,|\{p_i\})]^* F_{1,n}^{\Phi}(p\,|\{p_i\})\,e^{i(E_p-E)t} \right\}\,,
\label{firstorder}
\EN
we have
\EQ
D_\Phi=-\lambda\,B_\Phi(0)+O(\lambda^2)\,,
\label{D}
\EN
and
\EQ
\langle \Phi(x,t)\rangle_1=A_\Phi+\lambda\left[B_\Phi(t)-B_\Phi(0)\right]+O(\lambda^2)\,.
\label{onepoint2}
\EN

\section{Large time behavior}
\subsection{Perturbative results}
The time dependence of the one-point function (\ref{onepoint2}) is contained in the term (\ref{firstorder}). For $t\to\infty$ the integrand rapidly oscillates due to the factor $e^{i(E_p-E)t}$ and the leading contribution to the integral is obtained when all the momenta $p,p_1,\ldots,p_n$ are small. Notice that this is true also in case of cancellation of energy terms in the exponent. Indeed, $p$ is in any case made small by the large $R$ limit, so that $E_p\simeq M+p^2/2M$ can only be canceled by a single energy term $E_{p_i}$, with the consequence that $p_i=p$ is also small.

The matrix elements in (\ref{firstorder}) can be rewritten in terms of the form factors 
\EQ
F_n^{\cal O}(p_1,\ldots,p_n)=\langle 0|{\cal O}(0,0)|p_1,\ldots,p_n\rangle
\EN
by crossing the quasiparticle on the left. For $p,p_1,\dots,p_n\to0$ this gives
\begin{align}
F_{1,n}^{\cal O}(p|p_1,\ldots,p_n)&= F_{n+1}^{\cal O}(\bar{p},p_1,...,p_n)\nonumber\\
&+\sum_{i=1}^n 2\pi E_{p_i} \delta (p_i-p)\,(-1)^{i-1}\, F_{n-1}^{\cal O} (p_1,...,p_{i-1},p_{i+1},...,p_n)\,,
\label{matrix_element_1,n}
\end{align}
where $\bar{p}$ corresponds to the crossed quasiparticle with momentum $-p$ and energy $-E_p$, the first term on the r.h.s. is the connected part, and the terms with the delta function are the disconnected parts produced by the annihilation between the crossed quasiparticle and the quasiparticle with momentum $p_i$. In writing (\ref{matrix_element_1,n}) we took into account that for small momenta interacting theories in one spatial dimension exhibit fermionic statistics; this produces the factor $(-1)^{i-1}$ when the crossed quasiparticle reaches the quasiparticle to be annihilated. It follows that the product of matrix elements in (\ref{firstorder}) is given by
\begin{align}
&[F^{\Psi}_{1,n}(P|\{p_i\})]^* F^{\Phi}_{1,n}(p|\{p_i\})= [F^{\Psi}_{n+1}(\bar{P},p_1,...,p_n)]^* F^{\Phi}_{n+1}(\bar{p},p_1,...,p_n) \nonumber\\
&+ [F^{\Psi}_{n+1}(\bar{P},p_1,...,p_n)]^* \sum_{i=1}^n 2\pi E_{p_i} \delta (p_i-p)\,(-1)^{i-1}\, F_{n-1}^{\Phi} (p_1,...,p_{i-1},p_{i+1},...,p_n) \nonumber\\
&+ F^{\Phi}_{n+1}(\bar{p},p_1,...,p_n) \sum_{i=1}^n 2\pi E_{p_i} \delta (p_i-P)\,(-1)^{i-1}\, \left[F_{n-1}^{\Psi} (p_1,...,p_{i-1},p_{i+1},...,p_n)\right]^* \nonumber\\
&+ \sum_{i=1}^n \sum_{j=1}^n (2\pi)^2 E_{p_i} E_{p_j} \delta (p_i-P) \delta(p_j-p)\,(-1)^{i+j-2}\, \left[F_{n-1}^{\Psi} (p_1,...,p_{i-1},p_{i+1},...,p_n)\right]^*\nonumber\\
&\times F_{n-1}^{\Phi} (p_1,...,p_{j-1},p_{j+1},...,p_n)\,.
\label{product_matrix_elements}
\end{align}
We call the four terms in the r.h.s. connected-connected ($c_{\Psi}c_{\Phi}$), connected-disconnected ($c_{\Psi}d_{\Phi}$), disconnected-connected ($d_{\Psi}c_{\Phi}$) and disconnected-disconnected ($d_{\Psi}d_{\Phi}$), respectively. For the latter we further distinguish the terms with $i=j$ ($d^i_{\Psi}d^i_{\Phi}$) from those with $i\neq j$ ($d^i_{\Psi}d^j_{\Phi}$). Inserting (\ref{product_matrix_elements}) back into (\ref{firstorder}), the exponential factor in the integrand for each type of contribution reads
\begin{itemize}
\item $c_{\Psi}c_{\Phi}: e^{-\frac{R^2}{2}(p+P^2)+i[M(1-n)t-\frac{t}{2M}\sum_{i=1}^n p_i^2]}$
\item $c_{\Psi}d_{\Phi}: e^{-\frac{R^2}{2}(p_i+P^2)+i[M(1-n)t-\frac{t}{2M}\sum_{k\neq i}^n p_k^2]}$
\item $d_{\Psi}c_{\Phi}: e^{-\frac{R^2}{2}(p+p_i^2)+i[M(1-n)t-\frac{t}{2M}\sum_{k\neq i}^n p_k^2]}$
\item $d^i_{\Psi}d^i_{\Phi}: e^{-R^2 p_i^2+i[M(1-n)t-\frac{t}{2M}\sum_{k\neq i}^n p_k^2]}$
\item $d^i_{\Psi}d^j_{\Phi}: e^{-\frac{R^2}{2} (p_i^2+p_j^2)+i[M(1-n)t-\frac{t}{2M}\sum_{k\neq i,j}^n p_k^2]}$\,.
\end{itemize} 
Due to the delta functions, some energies can cancel in $e^{i(E_p-E)t}$ and no longer couple to $t$; as already observed, however, the corresponding momenta are still made small by the large $R$ limit. Moreover, if $p_i^2$ coming from $E_{p_i}\simeq M+p_i^2/2M$ couples in the exponential both to $R^2$ and $t$, we have $(R^2\pm it/M)p_i^2\to R^2p_i^2$ in the limit $R\to\infty$, which must be taken before that of large times.

For small momenta the form factors $F^{\Phi}_{n+1}$ in (\ref{product_matrix_elements}) behave as 
\EQ
\prod_{i=1}^n \frac{1}{p-p_i}\, \prod_{1\leq i<j\leq n} (p_i-p_j)\,,
\label{lowenergy}
\EN
where the numerator accounts for the fermionic statistics and the denominator for the annihilation poles\footnote{See \cite{Smirnov} for explicit illustrations in the case of integrable theories.}; the same is true for $F^{\Psi}_{n+1}$ after replacing $p$ with $P$. On the other hand, the form factors $F^{\Phi}_{n-1}$ and $F^{\Psi}_{n-1}$, which are the products of the annihilations, behave as $\prod_{k<l\, (k,l\neq i)}(p_k-p_l)$. The large time behavior of the different contributions to (\ref{onepoint2}) is now easily determined rescaling the momenta which couple to $t$ in the exponent. Up to the oscillating factor $e^{-i(n-1)Mt}$ we have
\begin{itemize}
\item $c_{\Psi}c_{\Phi}: t^{-n(n-1)/2}$
\item $c_{\Psi}d_{\Phi}, d_{\Psi}c_{\Phi}, d^i_{\Psi}d^i_{\Phi}: t^{-n(n-2)/2}$
\item $d^i_{\Psi}d^j_{\Phi}: t^{-(n-2)^2/2}$\,.
\end{itemize}
We see that the leading contribution at large time comes from $n=0$ ($c_{\Psi}c_{\Phi}$) and $n=2$ ($c_{\Psi}d_{\Phi}, d_{\Psi}c_{\Phi}, d^i_{\Psi}d^i_{\Phi}, d^i_{\Psi}d^j_{\Phi}$), and is purely oscillatory\footnote{We recall that $n\neq 1$ and that $n=0$ produces no disconnected part.}. It is easily checked in a similar way that the term $d^i_{\Psi}d^j_{\Phi}$ with $n=3$ actually vanishes as $R\to\infty$, so that the first subleading contribution at large $t$ comes from $c_{\Psi}c_{\Phi}$ with $n=2$ and is suppressed as $t^{-1}$.

Concerning the explicit calculation of the leading large time behavior of (\ref{firstorder}), it is straightforward for the $n=0$ contribution. For the $n=2$ contributions we know from (\ref{lowenergy}) that
\EQ
F_3^{\Phi}(\bar{p},p_1,p_2)\big\rvert_{p,p_1,p_2\to 0} \simeq a_{\Phi}\,\frac{p_1-p_2}{(p-p_1)(p-p_2)}\,,
\label{F3Phi}
\EN
and similarly
\EQ
F_3^{\Psi}(\bar{P},p_1,p_2)\big\rvert_{P,p_1,p_2\to 0} \simeq a_{\Psi}\,\frac{p_1-p_2}{p_1 p_2}\,.
\label{F3Psi}
\EN 
The fermionic statistics at low energies yields the expression
\begin{equation}
\underset{q_1=q_2}{\text{Res}}\, F_{k+2}^{\mathcal{O}}(\bar{q}_1,q_2,p_1,...,p_k)=iM \left[1-(-1)^k \right]\, F_k^{\mathcal{O}}(p_1,...,p_k)\,,
\label{residue}
\end{equation}
for the residue on an annihilation pole in the limit $q_1,q_2,p_1,...,p_k\to 0$ of our present interest. This in turn determines the coefficients
\begin{align}
&a_{\Phi}=2iM F_1^{\Phi}\,,\\
&a_{\Psi}=2iM F_1^{\Psi}\,.
\end{align}
The integrals in (\ref{firstorder}) are then computed using the expressions (\ref{F3Phi}) and (\ref{F3Psi}) with the prescription 
\EQ
\frac{1}{p-i\epsilon}=i\pi \delta(p)+\text{p.v.}\left(\frac{1}{p}\right)
\label{prescription}
\EN
for the poles. The results for the leading contributions are\footnote{They come only from the delta function terms in the pole prescription, since the principal values turn out to be subleading for $R\to\infty$.}
\begin{itemize}
\item $n=0$
\begin{itemize}
\item $c_{\Psi}c_{\Phi}: -\frac{2\sqrt{2}\lambda}{M^2}\, F_1^{\Psi} F_1^{\Phi}\, \cos Mt$
\end{itemize}
\item $n=2$
\begin{itemize}
\item $c_{\Psi}d_{\Phi}: \frac{2\lambda}{M^2} (1-\sqrt{2})\, F_1^{\Psi} F_1^{\Phi}\, \cos Mt$
\item $d_{\Psi}c_{\Phi}: \frac{2\lambda}{M^2} (1+\sqrt{2})\, F_1^{\Psi} F_1^{\Phi}\, \cos Mt$
\item $d^i_{\Psi}d^i_{\Phi}:  \frac{2\lambda}{M^2}\, F_1^{\Psi} F_1^{\Phi}\, \cos Mt$
\item $d^i_{\Psi}d^j_{\Phi}:  -\frac{2\sqrt{2}\lambda}{M^2}\, F_1^{\Psi} F_1^{\Phi}\, \cos Mt$\,,
\end{itemize}
\end{itemize}
where we took into account that for scalar Hermitian operators $F_1^{\cal O}$ is a real constant. Putting all together we obtain
\EQ
 B_\Phi(t)=\frac{F_1^{\Psi} F_1^{\Phi}}{M^2}\,(6-4\sqrt{2})\, \cos Mt+O(t^{-1})\,,\hspace{.7cm}t\to\infty\,.
 \label{Blarge}
 \EN 

We see from (\ref{onepoint2}) and (\ref{Blarge}) that for $t\to\infty$ the one-point function $\langle \Phi(x,t)\rangle_1$ tends to (if $F_1^{\Psi} F_1^{\Phi}=0$) or oscillates around (if $F_1^{\Psi} F_1^{\Phi}\neq 0$) the asymptotic offset $A_\Phi+D_\Phi$. The pre-quench value $A_\Phi$ is given by (\ref{prequench}) and involves $F_{1,1}^{\Phi}(p|q)$. Equations (\ref{matrix_element_1,n}) and (\ref{residue}) yield\footnote{It can be noted that, although the limit $R\to\infty$ in (\ref{prequench}) ensures that $p,q\to 0$, Eqs. (\ref{crossing2}) and (\ref{residue2}) hold for generic momenta. Indeed, the annihilation which produces the disconnected part requires no permutation, and no consideration on low energy statistics.} 
\EQ
F_{1,1}^{\cal O}(p|q)= F_{2}^{\cal O}(\bar{p},q)+2\pi E_{p}\,\delta(p-q)\, F_{0}^{\cal O}\,,
\label{crossing2}
\EN
\EQ
\underset{p=q}{\text{Res}}\, F_{2}^{\mathcal{O}}(\bar{p},q)=0\,.
\label{residue2}
\EN 
It follows that in the limit $R\to\infty$ implied in (\ref{prequench}) we obtain
\EQ
A_\Phi=F^\Phi_0=\langle 0|\Phi|0\rangle\,.
\label{prequench2}
\EN 
We see that the pre-quench value on the excited state coincides with that on the ground state\footnote{This result requires a generalization in the case of topological quasiparticles, to be discussed in the next section.}. On the other hand, since the post-quench state (\ref{psi1}) differs from that obtained from a quench from the ground state, the time evolution is in general different in the two cases. In particular, recalling also (\ref{D}), for the asymptotic offset we obtain
\EQ
E_\Phi\equiv A_\Phi+D_\Phi=\langle 0|\Phi|0\rangle-\lambda\,B_\Phi(0)+O(\lambda^2)\,,
\label{offset}
\EN
which differs at order $\lambda$ from that in (\ref{Phi_0}). Equations (\ref{Blarge}) and (\ref{offset}) lead to the large time result (\ref{Phi_1}).

\subsection{Further considerations}
\label{further}
Some additional considerations can be done starting from the perturbative result (\ref{Phi_1}). These go along the same lines applying to the quenches from the ground state, which we recall here. It was pointed out in \cite{quench} that the results of finite order perturbation theory can be expected to be quantitatively accurate up to a time scale $t_\lambda\sim 1/\lambda^{1/(2-X_\Psi)}$, where $X_\Psi$ is the scaling dimension of the quench operator. On the other hand, it was argued in \cite{oscill} that a first order result such as (\ref{Phi_0}) or (\ref{Phi_1}) with $F_1^{\Psi} F_1^{\Phi}\neq 0$ should lead to oscillations undamped beyond the perturbative scale $t_\lambda$. The argument focuses on the remainder of the perturbative series, namely the resummation of all terms beyond the first order. This will be a function that as time goes to infinity can either diverge, or approach a constant value, or itself exhibit undamped oscillations. Since the contribution of order $\lambda$ is bounded (recall (\ref{Phi_0}) and (\ref{Phi_1})), the first possibility is not expected for ordinary physical observables such as a local magnetization: at a given point of space this can grow in time to the limit of maximal ordering, but cannot diverge. Discarding then the possibility of divergence, the remaining two possibilities generically lead to undamped oscillations for the complete result (first order plus remainder) as long as these are present at first order\footnote{Of course the argument does not imply that the oscillations of the complete result are those of the first order.}. Hence, this argument of \cite{oscill} leads to the expectation of oscillations which can stay undamped beyond the perturbative time scale when $F_1^{\Psi} F_1^{\Phi}\neq 0$. Quite remarkably, oscillations exhibiting no damping over hundreds of periods within the accessible time interval were then observed numerically for a nonperturbative quench of the Ising spin chain in \cite{Jacopo}. This quench possesses in its analytical formulation only two time scales, the inverse mass gap $1/M$ and $t_\lambda$, and both are exceeded by {\it three orders of magnitude} in the nonperturbative case simulated in \cite{Jacopo}. The quench satisfies $F_1^{\Psi} F_1^{\Phi}\neq 0$, and it seems difficult to consider this full agreement between theoretical predictions and numerics as a coincidence\footnote{In this case one has to imagine a mechanism for generating some extremely late time scale beyond which the oscillations will eventually decay (see the discussion in \cite{BBK}). The decay of the oscillations of \cite{Jacopo} has not been observed to this date.}. The numerical, and also experimental \cite{Lukin}, observation of persistent oscillations has stimulated interest about ergodicity breaking. In this context, it has also been proposed that special symmetries can be related to the persistence of the oscillations (see e.g. \cite{MBJ} and, more generally, the references therein\footnote{On a more mathematical side, see also \cite{Buca}, where an attempt is made to describe the conditions under which ergodicity breaking may be expected via the concepts of pseudo-local operators and dynamical symmetries.}). While this role of symmetries does not emerge in our general framework, interesting intuition comes from the study of inhomogeneous quenches in \cite{oscillD}, where it is shown that, even for $F_1^{\Psi} F_1^{\Phi}\neq 0$, the oscillations decay already at first order in $\lambda$ if the quench is limited to a domain which is not extensive in all the spatial dimensions occupied by the system. In this case the energy density injected through the quench goes to zero at large times in any point of space, and is insufficient to sustain the oscillations.

It is also interesting to ask whether oscillations undamped for all times can be seen in an exact analytical solution. The possibility of exactly solving the quench (\ref{quench}) was addressed in \cite{quench} in the field theoretical framework, which provides a general definition of integrability at equilibrium (factorized scattering \cite{ZZ}). It was shown there that the quench (\ref{quench}) cannot be exactly solved in the case of interacting quasiparticles, the only one which can support undamped oscillations ($F_1^\Psi\neq 0$). This result is expected to be very restrictive also for nonrelativistic theories, which can in principle be thought as limits of the relativistic ones, and this makes sense of the extreme difficulty registered over the years also in the search of exact nonrelativistic solutions\footnote{See the case of \cite{DWBC}, which however does not fulfill the requirement for undamped oscillations.}. 

In the literature\footnote{The discussion of this point was solicited by an anonymous referee.}, the quest for exact treatment suggests sometimes to relax the definition (\ref{quench}) and to extend the name "quench" to the problem of the unitary evolution for $t>0$ of an assigned initial state, with no notion of a pre-quench state. The idea is that a suitably chosen initial state can mimic what happens in the genuine quench (\ref{quench}). For historical reasons inherited from the free fermion case of \cite{BMcD}, and of course for technical simplicity, the class of initial states usually considered in this context is that made of pairs of quasiparticles with opposite momenta ("pair structure"). The shortcoming of this choice is that, as shown in \cite{quench} for quenches from the ground state and again in (\ref{psi1}) in the present case, a huge simplification such as the pair structure does not arise in the quench (\ref{quench}) if the quasiparticles interact\footnote{See \cite{lightcone} for the relation between the structure of the nonequilibrium state and the light cone spreading of correlations. Implementations by hand of the pair structure in the context of specifically engineered lattice Hamiltonians have been presented in \cite{PPV}.}. The time evolution of a state with the pair structure has been considered in \cite{CS} in order to mimic a quench in the attractive regime of the sine-Gordon model. Besides the pair structure, the authors of \cite{CS} assumed also its exponentiation\footnote{These assumptions (see also \cite{GDLP}) were inspired by the integrable boundary states of equilibrium theories studied in \cite{GZ}.}. Expanding the exponential in powers and then resumming for the one-point functions they found oscillations exponentially suppressed in time. It is clear that this procedure has no impact on the aforementioned considerations about the perturbative expansion in $\lambda$. In \cite{CS} the quench (\ref{quench}) is not performed, there is no parameter $\lambda$, and no comparison is possible. The exponential decay of \cite{CS} is found assuming a state with an exponentiated pair structure: as we saw, both the pair structure and, {\it a fortiori}, its exponentiation are absent in the quench (\ref{quench}) with interacting quasiparticles. Since the state assumed in \cite{CS} does not belong to the set allowed for interacting quasiparticles, it cannot yield information about the dynamics of the sine-Gordon model following the quench (\ref{quench}). A quantitative study of a quench from the ground state in sine-Gordon has been performed in \cite{C-AH} in the perturbative framework. This quench has been studied numerically in \cite{EK}, where the persistent oscillations have been observed with frequencies related to the breather masses, in full agreement with the theoretical analysis of \cite{DV,C-AH}.

\section{Topological quasiparticles}
Some additional considerations are needed if the quench is performed within a phase with spontaneously broken symmetry. In the one-dimensional case we are considering this means that there are degenerate ground states $|0_a\rangle$ labeled by $a=1,2,\ldots,N$, and that the fundamental quasiparticle excitations have a topological nature, namely they are kinks $|K_{ab}(q)\rangle$ interpolating between $|0_a\rangle$ and $|0_b\rangle$, with $a\neq b$. It follows that the first excited state we consider in this paper as the pre-quench state now corresponds to such a kink. Then, when considering the pre-quench expectation value (\ref{prequench}), Eq.~(\ref{crossing2}) still holds with 
\EQ
F_{0}^{\cal O}=\langle 0_a|{\cal O}(0,0)|0_a\rangle\equiv\langle {\cal O}\rangle_a\,,
\EN
and
\EQ
F_{2}^{\cal O}(\bar{p},q)=\langle 0_a|{\cal O}(0,0)|K_{ab}(\bar{p})K_{ba}(q)\rangle\,.
\EN
Now, however, (\ref{residue2}) is replaced by\footnote{While (\ref{residue3}) was considered in \cite{DC_98} in the context of an integrable theory, the derivation given there is general, since involves no scattering.} \cite{DC_98}
\EQ
\underset{p=q}{\text{Res}}\, F_{2}^{\cal O}(\bar{p},q)=iM[\langle{\cal O}\rangle_a-\langle{\cal O}\rangle_b]\,.
\label{residue3}
\EN 
The pole associated to (\ref{residue3}) now gives an additional contribution when (\ref{prequench}) is evaluated using (\ref{prescription}), with the result that the pre-quench expectation value becomes
\EQ
A_\Phi=\frac{\langle\Phi\rangle_a+\langle\Phi\rangle_b}{2}\,.
\label{prequench3}
\EN 
The meaning of this result is quite clear. In the kink state, the system is in the ground state $|0_a\rangle$ on one side of the spatial location of the kink, and in the ground state $|0_b\rangle$ on the other side. Since the kink has a definite momentum $q$, it is completely delocalized in space, so that the expectation value of the field is given by the average (\ref{prequench3}). 

With this new expression for $A_\Phi$, the post-quench one-point function is still given by
(\ref{onepoint2}) and (\ref{firstorder}). Concerning its large time behavior, an undamped oscillating term would be again proportional to $F_1^{\Psi} F_1^{\Phi}$. However, the physically relevant cases correspond to a topologically neutral quench operator $\Psi$, and this implies that $F_1^{\Psi}=\langle 0_a|\Psi(0,0)|K_{ba}(q)\rangle$ vanishes. It follows that the large time result (\ref{Phi_1}) now becomes
\EQ
\langle \Phi(x,t)\rangle_1 =\frac{\langle\Phi\rangle_a+\langle\Phi\rangle_b}{2}-\lambda\left[B_\Phi(0)+O(t^{-1})\right]+O(\lambda^2)\,.
\label{Phi_1kink}
\EN

It is interesting to compare this result with that obtained in \cite{q_int}, where the time evolution in a nonequilibrium state interpolating between the degenerate ground states $|0_a\rangle$ and $|0_b\rangle$ was studied in a "no-quench" setting. We mean by this that, instead of considering the nonequilibrium state produced by a quench as in the present paper, in \cite{q_int} an infinite-dimensional space of interpolating nonequilibrium states was considered. Hence, the two results for the one-point function at large time, while referring to the same topological setting, have a significantly different origin. It is then nontrivial that they turn out to exhibit similar features\footnote{For the case of \cite{q_int} we refer to $t\to\infty$ with $x$ fixed.}. In both cases undamped oscillations are absent and the approach to the asymptotic offset is through terms decaying as $t^{-1}$. In both cases the offset involves the average (\ref{prequench3}): in (\ref{Phi_1kink}) the expectation values $\langle\Phi\rangle_a$ and $\langle\Phi\rangle_b$ refer to the $t<0$ Hamiltonian, and there is the correction $-\lambda B_\Phi(0)$ due to the quench; in \cite{q_int} the offset is (\ref{prequench3}) with $\langle\Phi\rangle_a$ and $\langle\Phi\rangle_b$ referring to the $t>0$ Hamiltonian.

\section{Conclusion}
In this paper we studied analytically the dependence on the initial condition of the late time dynamics following a quantum quench of a generic homogeneous one-dimensional system. More precisely, we considered the case in which before the quench the system is in the first excited state of its Hamiltonian $H_0$, at variance with the case of quenches from the ground state for which analytical results had so far been available. We overcame the technical difficulties related to the non-normalizability of the excited state working in presence of a regulator which, once removed at the end of the calculations, leaves a finite result for the observables. In this way we showed, in particular, that the condition for the presence of persistent oscillations of one-point functions is not affected by the change of the initial condition and remains that first found in \cite{quench} for quenches from the ground state: persistent oscillations of one-point functions arise if the post-quench spectrum of excitations includes neutral quasiparticles, and if the observable couples to these quasiparticles. The argument of \cite{oscill} pointing to oscillations persisting beyond the perturbative scale of our calculations -- a circumstance indeed numerically observed in \cite{Jacopo} -- applies also to the present case. At the same time, the comparison with the case of quenches from the ground state also shows quantitative differences in the amplitude of the oscillations, as well as in the value of the asymptotic offset. 

Another difference with the case of quenches from the ground state is that those from the excited state heavily involve, already at first order in the quench parameter $\lambda$, the connectedness structure of the matrix elements, with disconnected parts playing a substantial role. In perspective, it will be of interest to apply the approach of this paper to the case of more excited pre-quench states, namely pre-quench states with a larger number of quasiparticles. It seems reasonable to expect that the condition for the presence of undamped oscillations at large times will continue to be $F_1^{\Psi} F_1^{\Phi}\neq 0$, and it will be interesting to check that this indeed emerges from the explicit calculations. The role of disconnected parts in nonequilibrium quantum dynamics was already pointed out in \cite{lightcone}, where it was generally shown that they are responsible for the light cone propagation of two-point correlations. 

We also illustrated the implications of our results for quenches performed within a spontaneously broken phase, for which the first excited state corresponds to a topological excitation: a kink, or domain wall. We observed how the results for one-point functions at late times share quantitative properties with those obtained in \cite{q_int} in a very different way, namely considering an infinite-dimensional space of initial conditions of domain wall type, not necessarily produced by a quench.

Finally, it is worth emphasizing that the theory of quantum dynamics for extended systems has to be derived in the relativistic framework. The first reason is that it must include the vicinity of critical points. More generally, as we saw, also in the limit of large times ($t\gg 1/M$), where small momenta and nonrelativistic kinematics dominate, the connectedness structure of the relativistic theory plays an essential role. From this point of view, the situation is analogous to that exhibited by the fundamental theory of interfacial phenomena \cite{DV_interface,DS_intermediate,DSS_3d,DSS_wall}, where the linear size of the interface much larger than the bulk correlation length ($L\gg 1/M$) leads to the dominance of low energies, but the relativistic nature of the underlying theory remains determinant in the derivation of the main properties, including geometric effects \cite{wedge} and long range correlations \cite{DS_longrange,ST}.


\begin{thebibliography}{99}
\bibitem{q_int} G. Delfino and M. Sorba, Nucl. Phys. B 983 (2022) 115910.
\bibitem{BMcD} E. Barouch, B.M. McCoy and M. Dresden, Phys. Rev. A 2 (1970) 1075.
\bibitem{SPS} K. Sengupta, S. Powell and S. Sachdev, Phys. Rev. A 69 (2004) 053616.
\bibitem{CC} P. Calabrese and J. Cardy,  J. Stat. Mech. (2005) P04010.
\bibitem{quench} G. Delfino, J. Phys. A 47 (2014) 402001.
\bibitem{DV} G. Delfino and J. Viti, J. Phys. A: Math. Theor. 50 (2017) 084004.
\bibitem{oscill} G. Delfino, Nucl. Phys. B 954 (2020) 115002.
\bibitem{BCH} M.C. Banuls, J.I. Cirac and M.B. Hastings, Phys. Rev. Lett. 106 (2011) 050405.
\bibitem{RMCKT} T. Rakovszky, M. Mestyan, M. Collura, M. Kormos and G. Takacs,  Nucl. Phys. B 911, 805 (2016).
\bibitem{KCTC} M. Kormos, M. Collura, G. Takacs, and P. Calabrese, Nat. Phys. 13, 246 (2017).
\bibitem{Lukin} H. Bernien, S. Schwartz, A. Keesling, H. Levine, A. Omran, H. Pichler, S. Choi, A.S. Zibrov, M. Endres, M. Greiner, V. Vuletic and M.D. Lukin, Nature 551, 579 (2017).
\bibitem{Liu} F. Liu, R. Lundgren, P. Titum, G. Pagano, J. Zhang, C. Monroe and A.V. Gorshkov, Phys. Rev. Lett. 122, 150601 (2019).
\bibitem{Jacopo} O.A. Castro-Alvaredo, M. Lencs\'es, I.M. Sz\'ecs\'enyi and J. Viti, Phys. Rev. Lett. 124 (2020) 230601.
\bibitem{oscillD} G. Delfino and M. Sorba, Nucl. Phys. B 974 (2022) 115643.
\bibitem{nonint} G. Delfino, G. Mussardo and P. Simonetti, Nucl. Phys. B 473 (1996) 469.
\bibitem{Smirnov} F.A. Smirnov, Form factors in completely integrable models of quantum field theory, World
Scientific, 1992.
\bibitem{DC_98} G. Delfino and J. Cardy, Nucl. Phys. B 519 (1998) 551.
\bibitem{BBK} S. Birnkammer, A. Bastianello and M. Knap, Nature Communications 13 (2022) 7663.
\bibitem{MBJ} M. Medenjak, B. Buca and D. Jaksch, Phys. Rev. B 102 (2020) 041117.
\bibitem{Buca} B. Buca, Unified theory of local quantum many-body dynamics: Eigenoperator thermalization theorems, arXiv:2301.07091
\bibitem{ZZ} A.B. Zamolodchikov and Al.B. Zamolodchikov, Ann. Phys. 120 (1979) 253.
\bibitem{DWBC} J. De Nardis, B. Wouters, M. Brockmann and J.-S. Caux, Physical
Review A 89 (2014) 033601.
\bibitem{lightcone} G. Delfino, Phys. Rev. E 97 (2018) 062138.
\bibitem{PPV} L. Piroli, B. Pozsgay and E. Vernier, Nucl. Phys. B 925 (2017) 362.
\bibitem{CS} A. Cortes Cubero and D. Schuricht, J. Stat. Mech. 10 (2017) 103106. 
\bibitem{GDLP} V. Gritsev, E. Demler, M. Lukin, A. Polkovnikov, Phys. Rev. Lett. 99 (2007) 200404.
\bibitem{GZ} S. Ghoshal and A.B. Zamolodchikov, Int. J. Mod. Phys. A9 (1994) 3841.
\bibitem{C-AH} O. A. Castro-Alvaredo and D.X.  Horv\'ath, SciPost Phys. 10, 132 (2021).
\bibitem{EK} P. Emonts and  I. Kukuljan, Phys. Rev. Res. 4 (2022) 033039.
\bibitem{DV_interface}  G. Delfino and J. Viti, J. Stat. Mech. (2012) 10009.
\bibitem{DS_intermediate} G. Delfino and A. Squarcini, Annals of Physics 342 (2014) 171.
\bibitem{DSS_3d} G. Delfino, W. Selke and A. Squarcini, Nucl. Phys. B 958 (2020) 115139.
\bibitem{DSS_wall} G. Delfino, M. Sorba and A. Squarcini, Nucl. Phys. B 967 (2021) 115396.
\bibitem{wedge} G. Delfino and A. Squarcini, Phys. Rev. Lett. 113 (2014) 066101.
\bibitem{DS_longrange} G. Delfino and A. Squarcini, JHEP 11 (2016) 119.
\bibitem{ST} A. Squarcini and A. Tinti, JHEP 03 (2023) 123.



\end{thebibliography}
\end{document}